\documentclass[11pt,a4paper]{article}
\usepackage{amsfonts,amssymb,amsmath,amsthm,cite}
\usepackage{graphicx}
\usepackage{slashed}
\usepackage{hyperref}

\evensidemargin=\oddsidemargin
\newcommand{\MM}{\mathcal{M}}
\newcommand{\ii}{\mathrm{i}}
\newcommand{\nn}{\mathbf{n}}

\newcommand{\dd}{\mathrm{d}}

\begin{document}
\title{Edge states and the $\eta$ invariant}
\author{Rodrigo Fresneda\thanks{Email: rodrigo.fresneda@ufabc.edu.br}, Lucas de Souza\thanks{Email: lucas.souza4799@gmail.com} and Dmitri Vassilevich\thanks{Email: dvassil@gmail.com}\\Center of Mathematics, Computation and Cognition, Universidade Federal do ABC\\
 09210-580, Santo Andr\'e, SP, Brazil}
\date{\empty}
\maketitle

\begin{abstract}
We propose a relation between the $\eta$ invariant on a manifold with boundary, the $\eta$ invariants of edge states, and the $\eta$ invariant in an infinite volume limit. With the example of planar fermions with bag and chiral bag boundary conditions we show that this relation holds whenever edge states are sufficiently well-localized near the boundary. As a by-product we show that the spectrum of edge modes for chiral bag boundary conditions is linear but bounded.
\end{abstract}

\section{Introduction}\label{sec:Intro}
The $\eta$ invariant plays an important role in Quantum Field Theory. It describes the parity anomaly \cite{Redlich:1983dv,Niemi:1983rq,AlvarezGaume:1984nf} which leads to a Chern--Simons term in the effective action for gauge fields and to a Hall type conductivity in condensed matter applications. The Jackiw--Rebbi fermion number fractionization \cite{Jackiw:1975fn} can be also described through the $\eta$ invariant, see \cite{Niemi:1983rq,Niemi:1984vz}. 

If there are boundaries, the $\eta$ invariant has specific boundary contributions. For example, on four-dimensional manifolds these contributions take a form of Chern--Simons actions on the three-dimensional boundary manifold \cite{Kurkov:2017cdz,Kurkov:2018pjw}. Much attention has been paid to the relations between $\eta$ invariants in the bulk and $\eta$ invariants in effective boundary theories. Important results in this direction were obtained \cite{Witten:2019bou} in the context of anomaly inflow,  where an expression for $\eta$ in terms of boundary states was derived. However, many problems still remain open. The boundary conditions used in \cite{Witten:2019bou} (and earlier in \cite{Witten:2015aba}) are non-hermitian and thus cannot be used for the Dirac Hamiltonians and for the problem of fermion number fractionization. In \cite{Ivanov:2022jor}, the $\eta$ invariant for local boundary conditions leading to a hermitian Dirac operator were expressed through the anomalies in effective boundary theories. However, only massless Dirac operators do not admitting any boundary states were considered in the work \cite{Ivanov:2022jor}. (For completeness, we also mention a work \cite{Fukaya:2020tjk} which considered domain walls rather than boundaries.)

The purpose of this work is to derive the relations between the $\eta$ invariant on a manifold with boundary and the boundary (edge) modes in theories with hermitian Dirac operators.

Let us give some main definitions. For more details we refer the reader to \cite{GilkeyNew}.
Let $H$ be a hermitian Dirac type operator on a smooth manifold $\MM$, $\dim\MM=n$, with a smooth boundary $\partial\MM$ with eigenvalues $\lambda$. The $\eta$ function of this operator is defined as
\begin{equation}
\eta(s,H)=\sum_{\lambda>0}\lambda^{-s}-\sum_{\lambda<0} (-\lambda)^{-s},
\label{eta}
\end{equation}
where $s$ is a complex parameter. The sum in (\ref{eta}) is convergent if $\Re s$ is large enough and defines a meromorphic function in the whole complex plane. The $\eta(0,H)$ measures the spectral asymmetry of $H$. Note that zero eigenvalues of $H$ are not included in (\ref{eta}). The Atiyah--Patodi--Singer $\eta$ invariant is defined as
$
\eta_H=\tfrac 12 (\eta(0,H)+\dim\, \mathrm{Ker}(H)) $.
This invariant jumps by $\pm 1$ whenever an eigenvalue crosses the origin. The exponentiated $\eta$ invariant $
\mathcal{E}(H)=\exp (-2\pi \ii \eta_H)$
is smooth.

Let $t$ be a positive real parameter. Then, there is an asymptotic expansion at $t\to +0$ of the heat kernel
\begin{equation}
\mathrm{Tr}\, \left( Q \exp(-tH^2)\right) = \sum_{k=0}^\infty t^{\frac{k-n}{2}}a_k(Q,H^2)\label{hkexp}
\end{equation}
for any smooth matrix-valued function $Q$. The heat kernel coefficients are very well studied in the literature, see e.g. \cite{Vassilevich:2003xt}. Let $\delta H$ be a variation of $H$. We assume that $\delta H$ is a matrix-valued function rather than a differential operator. If due to this variation no eigenvalue passes through $0$, the corresponding variation of the $\eta(0,H)$ reads \cite{AlvarezGaume:1984nf,Atiyah:1980jh,Gilkey:1984}
\begin{equation}
\delta\eta(0,H)=-\frac 2{\sqrt{\pi}} a_{n-1}(\delta H,H^2). \label{vareta}
\end{equation}
This formula will be our main technical tool. We will study exclusively the perturbative (smooth) contributions to $\eta(0,H)$. This is equivalent to saying that we will be interested in the logarithm of the exponentiated $\eta$ invariant.

The heat kernel expansion and the formula (\ref{vareta}) are valid provided the boundary conditions are strongly elliptic. Without discussing this requirement in detail, we just mention that all boundary conditions used in this work are strongly elliptic. Let us also note that (\ref{vareta}) can be used for arbitrary smooth variations on compact manifolds while on noncompact manifolds the variations have to be localized and thus cannot change the asymptotic behaviour of background fields. 

We are going to argue that in some situations $\eta(0)$ can be represented as a sum of a bulk part and a contribution of boundary states. It is not easy to separate the bulk and boundary contributions in general, especially since in many cases $\eta(0)$ is topological and may be presented as a volume integral of a topological density as well as a surface integral. To make this separation transparent, we propose the following scheme. Consider a Dirac type operator on $\mathbb{R}^n$ such that all background fields entering this operator are localized near the origin. (Example: a gauge field with field strength rapidly decaying away of the origin). We introduce a spherical boundary $S^{n-1}$ of a radius $r$ and impose some boundary conditions. We expect that if $r$ is sufficiently large the $\eta$ invariant on the ball splits into two parts, one being the $\eta$ invariant for $\mathbb{R}^n$ without boundaries while the other part is given by contribution of the edge states. The edge states are the eigenstates of Dirac Hamiltonian which decay exponentially fast as functions of the distance from the boundary. The decay rate has to be sufficiently fast to ensure localization of the states in a small neighbourhood of the boundary.

In this paper, we check the proposed relation between $\eta$ functions with the simplest yet nontrivial example of planar fermions in an external magnetic field. In the next Section, we consider bag boundary conditions, we show that the edge states are localized near the boundary, and  we confirm the relation. In Section \ref{sec:more}, we generalize boundary conditions to the chiral bag boundary conditions. We demonstrate that the edge states have a non-Dirac spectrum which is linear but bounded. Besides, near the threshold the edge states are not localized leading to a violations of the relation between $\eta$ functions.

\section{Bag boundary conditions}\label{sec:bag}
Consider a plane $\mathbb{R}^2$ pierced by a perpendicular magnetic field. We assume that this field has a finite flux and is concentrated somewhere near the origin. The Dirac Hamiltonian for fermions confined to the plane reads
\begin{equation}
H=\alpha^j (\ii\partial_j - eA_j) - \beta m, \label{DirHam}
\end{equation}
where $\alpha^j$ and $\beta$ are hermitian $2\times 2$ matrices satisfying
\begin{equation}
\mathrm{tr}\, \left(\alpha^k\alpha^j\beta \right)=-2\ii \epsilon^{kj}, \qquad \alpha^j\alpha^k + \alpha^k\alpha^j = 2g^{jk}, \label{tr}
\end{equation}
with $\epsilon^{jk}$ and $g^{jk}$ being the Levi-Civita tensor and the metric tensor, respectively. Besides, $\beta\alpha^j+\alpha^j\beta=0$ and $\beta^2=1$.

The famous Niemi--Semenoff \cite{Niemi:1983rq} result for infinite plane is
\begin{equation}
\eta(0,H)=-\frac e{4\pi}\, \frac m{|m|} \int_{\mathbb{R}^2} \dd^2x\, \epsilon^{jk}F_{jk}\,. \label{NS}
\end{equation}

Let us consider a disk $D_r$ of radius $r$ centred at the origin. Let $\nn$ be a unit inward pointing normal to the boundary $\partial D_r=S^1_r$, $\theta$ be a coordinate on $S^1_r$, and $\sqrt{h}\dd\theta=r\dd\theta$ be the induced integration measure. We introduce
\begin{equation}
\eta_r(0,H):=\frac e{2\pi}\, \frac m{|m|} \int_{S_r^1} \dd\theta\, \sqrt{h}\, A_j \epsilon^{\nn j} \label{etar}
\end{equation}
so that
\begin{equation}
\eta(0,H)=\lim_{r\to\infty} \eta_r(0,H).\label{etalim}
\end{equation}
We stress that $\eta_r$ is not an $\eta$ function of any operator.

Let us compute the $\eta$ function for $H_{D_r}$ which is $H$ restricted to $D_r$ with boundary conditions
\begin{equation}
\Pi_-\psi\vert_{S^1_r}=0 \label{bc}
\end{equation}
with 
\begin{equation}
\Pi_\pm =\tfrac 12 (1 \pm \ii \varepsilon\beta\alpha^{\nn}),\label{Pibag}
\end{equation}
where $\varepsilon=\pm 1$. We also define
\begin{equation}
\chi:=\Pi_+ - \Pi_-,\qquad \Pi_\pm =\frac 12 (1\pm \chi).\label{chi}
\end{equation}
The boundary conditions (\ref{bc}), (\ref{Pibag}) are known under the name of bag boundary conditions \cite{Chodos:1974je,Chodos:1974pn} in the physics literature and under the name of Clifford boundary conditions in Mathematics \cite{Gilkey:1983a}. For these conditions, the normal component of fermion current, $\psi^\dag \alpha^\nn \psi$, vanishes at all point of the boundary, so that $H$ is selfadjoint. 

For any Dirac type operator ${H}$ on a two-dimensional manifold $\mathcal{M}$ with bag boundary conditions on $\partial\MM$
\begin{equation}
a_1(Q,H^2)=\frac 1{8\sqrt{\pi}} \int_{\partial \MM} \dd \theta\, \sqrt{h}\, \mathrm{tr}\, ( Q\chi ).\label{a1}
\end{equation}
By specifying $H=H_{D_r}$, $Q=\delta H_{D_r}=-e\alpha^j A_j$ and using the formula (\ref{vareta}) for $n=2$, we obtain
\begin{equation}
\delta\eta(0,H_{D_r})=\frac{\epsilon e}{2\pi}\int_{S^1_r} \dd\theta\, \sqrt{h}\, \delta A_j \epsilon^{\nn j},
\end{equation}
which is integrated to
\begin{equation}
\eta(0,H_{D_r})=\frac{\epsilon e}{2\pi}\int_{S^1_r} \dd\theta\, \sqrt{h}\, A_j \epsilon^{\nn j}. \label{etaDr}
\end{equation}

Let us analyse the edge states. We fix the gauge $A_\nn=0$ near the boundary and consider an auxiliary eigenvalue problem 
\begin{equation}
\hat{H}\varphi =\lambda \varphi \label{hatH}
\end{equation}
on $\mathbb{R}_+$ where the operator $\hat{H}$ is constructed from $H$ in the following way. Let $e^{\Vert}$ be a unit vector tangential to the boundary, so that $\bigl( \alpha^{\Vert}\bigr)^2=1$ with $\alpha^{\Vert}:=e_j^{\Vert}\alpha^j$. The corresponding coordinate on the circle is $x^{\Vert}=r\theta$. The function $\varphi$ is assumed to be an eigenfunction of $\ii \partial_{\Vert}-eA_\Vert$ with an eigenvalue $\xi$. At large $r$, the extrinsic curvature of the boundary is negligible. Thus, the eigenvalue problem (\ref{hatH}) reads
\begin{eqnarray}
\ii \alpha^\nn (\partial_\nn +q(\lambda,m,\xi))\varphi=0,\\
q(\lambda,m,\xi))=-\ii \alpha^\nn (\alpha^{\Vert}\xi -\beta m -\lambda)\,.
\end{eqnarray}
$\varphi$ depends on a single coordinate $x^\nn$ and satisfies the boundary condition 
\begin{equation}
\Pi_-\varphi(x^\nn=0)=0\,.
\end{equation}
Edge states correspond to 
\begin{eqnarray}
\varphi\to 0\quad \mbox{at}\quad x^\nn\to \infty ,
\end{eqnarray}
which means that the eigenspace of $q(\lambda,m,\xi)$ corresponding to a positive eigenvalue should coincide with the kernel of $\Pi_-$. To resolve this condition, let us take a particular representation of $\alpha^j$ and $\beta$,
\begin{equation}
\beta=\begin{pmatrix}
1 & 0 \\ 0 & -1
\end{pmatrix},\qquad
\alpha^\nn= \begin{pmatrix}
0 & 1 \\ 1 & 0
\end{pmatrix},\qquad
\alpha^{\Vert}=\begin{pmatrix}
0 & \ii \\ -\ii & 0
\end{pmatrix}.
\end{equation}
With this choice, $\epsilon^{\nn \Vert}=+1$.

We have,
\begin{equation}
q(\lambda, m, \xi)=\begin{pmatrix}
-\xi & \ii (\lambda-m) \\ \ii (\lambda+m) & \xi
\end{pmatrix}. \label{qlmx}
\end{equation}
The eigenvalues of this matrix are $\pm \mu$ with $\mu= \sqrt{\xi^2 - \lambda^2 + m^2}$ while
\begin{equation}
\begin{pmatrix}
-\xi + \sqrt{\xi^{2}-\lambda^{2}+m^{2}} \\ \ii(\lambda+m) 
\end{pmatrix}\label{eigen1}
\end{equation}
is an eigenvector corresponding to the positive eigenvalue of $q(\lambda, m, \xi)$.

The kernel of the projector
\begin{equation}
\Pi_-=\dfrac{1}{2}\begin{pmatrix}
1 & -\ii\varepsilon \\ \ii\varepsilon & 1    
\end{pmatrix} \label{Pmax}
\end{equation}
is spanned by the vector
\begin{equation}
\begin{pmatrix}
-1 \\ \ii\varepsilon
\end{pmatrix}.\label{eigenPi}
\end{equation}
The vectors (\ref{eigen1}) and (\ref{eigenPi}) define the same linear subspace iff
\begin{equation}
\dfrac{m}{|m|} = -\varepsilon \quad \mbox{and} \quad \lambda = \varepsilon\xi\,.\label{bstcon}
\end{equation}
The fist equation in (\ref{bstcon}) defines the boundary condition when edge state exist, while the second one gives the spectrum of Dirac Hamiltonian $H$ on these modes. Thus we conclude that the restriction of $H$ to edge states reads 
\begin{equation}
H_{\mathrm{b}}=\varepsilon(\ii \partial_{\Vert} -eA_\Vert) .\label{Hb}
\end{equation}

The dependence of edge states on $x^\nn$ is defined by the positive eigenvalue of $q$. Due to the second conditions in (\ref{bstcon}), this eigenvalue is $\mu=|m|$, so that whenever edge modes exist they decay as $e^{-|m|x^\nn}$ and are well localized relatively to $r$ as long as $|m|r$ is sufficiently large. 

The boundary Hamiltonian $H_{\mathrm{b}}$ is a one-dimensional Dirac type operator. The variation of $\eta(0,H_{\mathrm{b}})$ can be calculated by using (\ref{vareta}) for $n=1$,
\begin{eqnarray}
&&\delta\eta(0,H_{\mathrm{b}})=-\frac 2{\sqrt{\pi}} a_0(\delta H_{\mathrm{b}},H_{\mathrm{b}}^2)= -\frac 2{\sqrt{\pi}} \, \frac 1{\sqrt{4\pi}}
\int_{S^1_r} \dd x^\Vert\, \delta H_{\mathrm{b}}\nonumber\\
&&\qquad
=\frac{e\varepsilon}{\pi}\int_{S^1_r} \dd x^\Vert \delta A_\Vert = \frac{\varepsilon e}{\pi}\int_{S^1_r} \dd\theta \sqrt{h}\, \delta A_\theta \epsilon^{\nn\theta}.\label{delHb}
\end{eqnarray}

The variation (\ref{delHb}) is integrated to
\begin{equation}
\eta(0,H_{\mathrm{b}})=\frac{\varepsilon e}{\pi}\int_{S^1_r} \dd\theta \sqrt{h}\,  A_\theta \epsilon^{\nn\theta}. \label{etaHb}
\end{equation}

If $m/|m|=\varepsilon$ there are no boundary states and the expressions $\eta_r(0,H)$ coincides with $\eta(0,H_{D_r})$. If $m/|m|=-\varepsilon$ the two $\eta$ functions have opposite signs but their difference is compensated by the contribution of boundary modes (\ref{etaHb}). In general,
\begin{equation}
\eta(0,H_{D_r})=\eta_r (0,H)+\eta(0,H_{\mathrm{b}}). \label{etaetaeta}
\end{equation}
The second term on the right-hand side of (\ref{etaetaeta}) is zero if there are no boundary modes. At $r\to\infty$, this is exactly the relation which has been announced in the Introduction. We stress again that in (\ref{etaetaeta}) we relate smooth parts of the $\eta$ functions. In other words, our claim is that variations of both parts agree as long as eigenvalues do not cross the origin.

\section{More general boundary conditions}\label{sec:more}
A natural generalization of the boundary conditions considered in the previous Section consists in the modification of $\chi$ in (\ref{chi}) as
\begin{equation}
\chi =\ii \varepsilon \beta e^{\tau\beta}\alpha^\nn  ,\label{chibag}
\end{equation}
with $\tau$ being a real parameter. These conditions are the chiral bag boundary conditions \cite{Rho:1983bh} (modulo the replacement of a Dirac Hamiltonian by a Euclidean Dirac operator). For chiral bag boundary conditions, the normal component of fermion current vanishes and $H$ is selfadjoint. Moreover, these boundary conditions are strongly elliptic \cite{Beneventano:2003hv} so that the heat kernel expansion (\ref{hkexp}) exists and the formula (\ref{vareta}) for the variation of $\eta(0,H)$ is still valid. Although the relation (\ref{a1}) is not hold for $\chi$ given in (\ref{chibag}), it has been demonstrated \cite{Ivanov:2021yms} that the smooth part of
$\eta(0,H_{D_r})$ does not depend on $\tau$. Therefore, Eq.\ (\ref{etaDr}) can still be used. All what remains is to analyse the edge states. The calculations are similar to that of the previous Section, so that we do not give details here.

The edge states exist if
\begin{equation}
-\varepsilon (\lambda \sinh (\tau)+ m\cosh (\tau))\geq 0 \label{chiex}
\end{equation}
while the dispersion relation is given by
\begin{equation}
\lambda =\frac {\varepsilon\xi}{\cosh(\tau)} - m\tanh (\tau).\label{chidis}
\end{equation}
We see, that the spectrum of boundary Hamiltonian given by (\ref{chidis}) is linear, but edge states exist for any signs of $\varepsilon$ and $m$, while the spectrum is bounded either above or below. 

For Eq.\ (\ref{etaetaeta}) to hold, $\eta(0,H_b)$ must vanish when $m$ and $\varepsilon$ are of the same sign. We are not aware of any method of computations of $\eta$ invariants for operators with the spectrum restricted as in (\ref{chiex}). If we neglect this restriction for a while, $H_b=\varepsilon (\cosh(\tau))^{-1}(\ii \partial_{\Vert} -eA_\Vert) - m\tanh (\tau)$. This is an operator of Dirac type which can be treated along the same lines as the boundary Hamiltonian in the previous Section. One thus obtains a non-zero value of $\eta(0,H_b)$ which depends on $A_\Vert$ and $m$. It is hard to conceive that imposing the restriction (\ref{chiex}) (which make the spectrum even more asymmetric) can kill the spectral asymmetry. In fact, Eq.\ (\ref{etaetaeta}) does not need to hold for chiral bag boundary conditions. The positive eigenvalue of $q$ which defines the decay rate of edge states reads $|\lambda \sinh (\tau)+ m\cosh (\tau)|$. Near the threshold of the spectrum (\ref{chiex}) the decay rate is very small, so that the edge modes are \emph{not} well localized.

\section{Conclusions and discussion}\label{sec:con}
In this paper, we have shown with two examples that $\eta(0,H)$ can be represented as a sum of a bulk part and a contribution from the edge states whenever the latter ones are localized in a sufficiently small neighbourhood of the boundary. This gives a new interpretation of the classical result \cite{Niemi:1983rq} and opens an avenue for future research. A rigorous demonstration of this relation for generic Dirac operators is a complicated problem in Mathematics, which we are going to address in the near future.

As a by-product we derived the spectrum of edge modes for chiral bag boundary conditions. This spectrum reminds us of the Fermi arc boundary states in Weyl semimetals  \cite{Armitage:2017cjs} which also occupy a bounded region in the momentum space and points out to new physical applications.

Our results may be applied to calculations of fermion fractionization in the presence of solitons. This quantity is usually a function of asymptotic values of background fields (though there is an exception \cite{Almeida:2021lks}) so that the formula (\ref{vareta}) is not applicable. Having an expression for charge fractionization through $\eta$ functions on compact manifolds may simplify the calculations.

Finally, we would like to mention some publications which treated spectral properties and anomalies of Dirac operator and with bag \cite{Hortacsu:1980kv,Hrasko:1983sj,Wipf:1994dy,Kirchberg:2006wu}, chiral bag \cite{Gilkey:2005qm} and even some other boundary conditions \cite{Sitenko:2014kza,Angelone:2022yuu}. The setup used in these papers was quite different from ours.

\paragraph{Acknowledgements.} This work was supported in parts by the S\~ao Paulo Research Foundation (FAPESP), grant 2021/10128-0. Besides, the work of LS was supported by the grant 2022/09068-6 of FAPESP, and DV was supported by the National Council for Scientific and Technological Development (CNPq), grant 304758/2022-1.

\bibliographystyle{fullsort}
\bibliography{parity}
\end{document}